\documentclass[letterpaper,twocolumn,10pt]{article}
\usepackage{usenix,epsfig,endnotes,url}

\usepackage{graphicx}
\usepackage{subfigure}
\usepackage{algpseudocode}
\usepackage{algorithm}
\usepackage{amsmath}
\usepackage{multirow}
\usepackage{float}
\usepackage{amsmath}
\usepackage{amssymb}

\begin{document}

\date{}

\title{\Large \bf Improving Raw Image Storage Efficiency by Exploiting Similarity}


\author{
{\rm Binqi Zhang, Chen Wang*, Bing Bing Zhou, Albert Y. Zomaya}\\
The University of Sydney\\
*CSIRO
}

\maketitle

\thispagestyle{empty}

\subsection*{Abstract}
To improve the temporal and spatial storage efficiency, researchers have intensively studied various techniques, including compression and deduplication. Through our evaluation, we find that methods such as photo tags or local features help to identify the content-based similarity between raw images. The images can then be compressed more efficiently to get better storage space savings. Furthermore, storing similar raw images together enables rapid data sorting, searching and retrieval if the images are stored in a distributed and large-scale environment by reducing fragmentation. In this paper, we evaluated the compressibility by designing experiments and observing the results.  We found that on a statistical basis the higher similarity photos have, the better compression results are. This research helps provide a clue for future large-scale storage system design.

\section{Introduction}

With the rapid growth of data volume, the efforts to optimize spatial and temporal efficiency have never stopped. Compression and deduplication are two well-known technologies to save storage space. Studies ~\cite{aronovich2009design, yokoo1997data} find that applying deduplication and compression techniques on similar data helps achieve better results. On the other hand, as more data are stored in a distributed environment because of the scale, the placement of data becomes important. If similar data are placed on the same node, or even a smaller number of nodes, the read performance can be significantly better than a highly fragmented placement. In a backup system, the reduction of fragmentation helps improve the performance of data restore ~\cite{fu2015design}. Therefore, for large-scale storage system, the benefit of using similarity to determine data placement is twofold: first, it helps deduplication or compression save more storage space; second, it enables quick search, sorting and read operations. 

In addition, researchers find general compression or deduplication methods may not work well for all workloads and data sets. In the recent years, workload-aware deduplication or compression techniques have been proposed ~\cite{lin2015metadata, dewakar2015storage}. Instead of just checking bit-wise similarity, examining contents to put data into similar groups, can be helpful to improve the storage efficiency more significantly. The program in ~\cite{shi2014photo} has used local features detection to help compress photos albums sharing many similar contents.

One of the most common use cases for cloud storage is to upload and share personal digital photos, via social media or image repository. Photos uploaded by one user, often by albums, are more likely to be similar in contents. Raw images are increasingly popular among professional photographers, photo hobbyists, healthcare IT professionals and scientific researchers. However, more efforts to optimize the storage efficiency for raw images that preserve visual similarity are needed, which may be complementary to JPEG encoding and compression.

To this end, we propose exploiting the detection of content-based similarity for raw images. The similarity should be utilized for better spatial and temporal storage efficiency. We present our observations and insights from two approaches (one based on photo tags, the other on local feature extraction) in exploring the compressibility. We have not intended to create a specific storage system design here. Instead, we would like to share our findings and inspire more work to substantiate the methods and optimize the performance for real-life raw image workloads. 

The main contributions of this paper are: 
\begin{itemize}
	\setlength{\itemsep}{1pt}
	\item We set up and perform empirical studies on two content-based similarity detection approaches to compress similar raw images.
	\item We analyze the results with statistical views and gain insights for future design.
	\item We discuss technical limitations and challenges. 
\end{itemize}

\section{Background}

To achieve better compressibility, LZMA used in 7z ~\cite{pavlov20137zip} employs larger sliding window. The compression program finds redundant strings within a certain length of window. With a greater window size, the chance of hitting redundant strings are bigger, thus the compression results are better. Similarly, rzip ~\cite{kolivas2008long} looks for identical contents over a longer distance throughout the file. It uses hash values for fixed size chunks for the check and this method allows better intra-file deduplication. 

These techniques do not guarantee that good storage efficiency is obtained at system level unless the system can feed the compression tools with the right set of data: in our case, the similar photos. Digital images are expressed in pixel values composing of basic colors such as red, green, blue, often denoted as R,G,B values respectively. In this study, we aim to analyze the storage efficiency for raw images. All images are presented in a set of pixels with R,G,B values. When speaking of ``similarity'' of photos, we may refer to the color, the pattern, the content or even the theme. Things get quite complicated. Some objects that human views as similar are regarded as totally different by computer because their binary values are not equal. For instance, two images with same pattern: one in red and the other in blue. The red one denotes (1, 0, 0) for all pixels while the blue one denotes (0, 0, 1). 

We do not focus on pixel level similarity detection as it is finer-grained and too complex. And it may not be viable as a pre-precessing step just to feed the compression tools for its costly computing. Then, we ask ourselves whether there is a good way to identify the basic similarity of the digital photos. Local features have been used to distinct two images. Methods such as SIFT \cite{lowe2004distinctive} have demonstrated invariance to scale or rotation and have been widely used in image processing.

\section{Approach for Empirical Study}

Today, when users post their photos to social media sites, they often mark the photo with text description, short expressions, classify the photos / album with keyword tags, such as ``sydney opera house'' or ``trip to sydney''. We assume such tagging mechanism, together with the data processing performed by the social media platform, helps to quickly locate similar contents for digital photos. The similarity, no matter in color, pattern, theme or a combination of them, should contribute to better compressibility using the aforementioned compression tools. We perform some evaluation work to assess the correlation between tags, similarity and compressibility. 

To validate our hypothesis, we design a set of experiments on publicly available data sets to exploit the relationship of image similarity and the compression ratio of associated photo groups. We use an open source application programming interface (API) Flickr4Java ~\cite{flickr4java} to download photos from Flickr. To reduce the number of photos that are not relevant to the tags, we choose ``relevance'' as the sorting method. The results returned by the Flickr platform are sorted by the API in the descent mode by relevance to the tag theme.

\begin{figure}[h!]
\centering
 \centerline{\includegraphics[width=0.5\textwidth, height=0.75in]{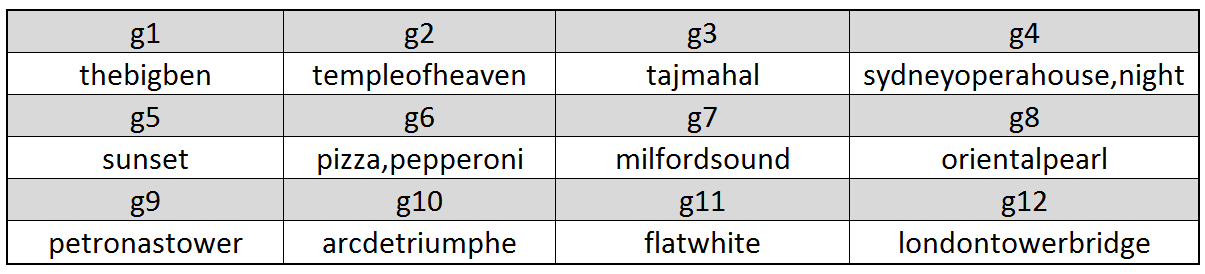}}
  	
   	\caption{Tag selection for photo groups}
\label{fig:tag}
\end{figure}

We select twelve tags for photo search and create photo groups according to their tags as listed in Fig.~\ref{fig:tag}. We also attempt to use multiple tags. The tags are delimited by a comma. When available, photos in original size are downloaded. If not, large, medium or small images are downloaded. So the size of the images vary, depending on the download authorization levels set by the image owners. We create subgroups consisting of 100, 50 and 20 most relevant photos for each tag/group respectively. And we create a comparison subgroup for SIFT-picked photos from the Top-100 one. The method is explained later in this section.

\begin{figure}[h!]
\centering
 \centerline{\includegraphics[width=0.5\textwidth, height=1.5in]{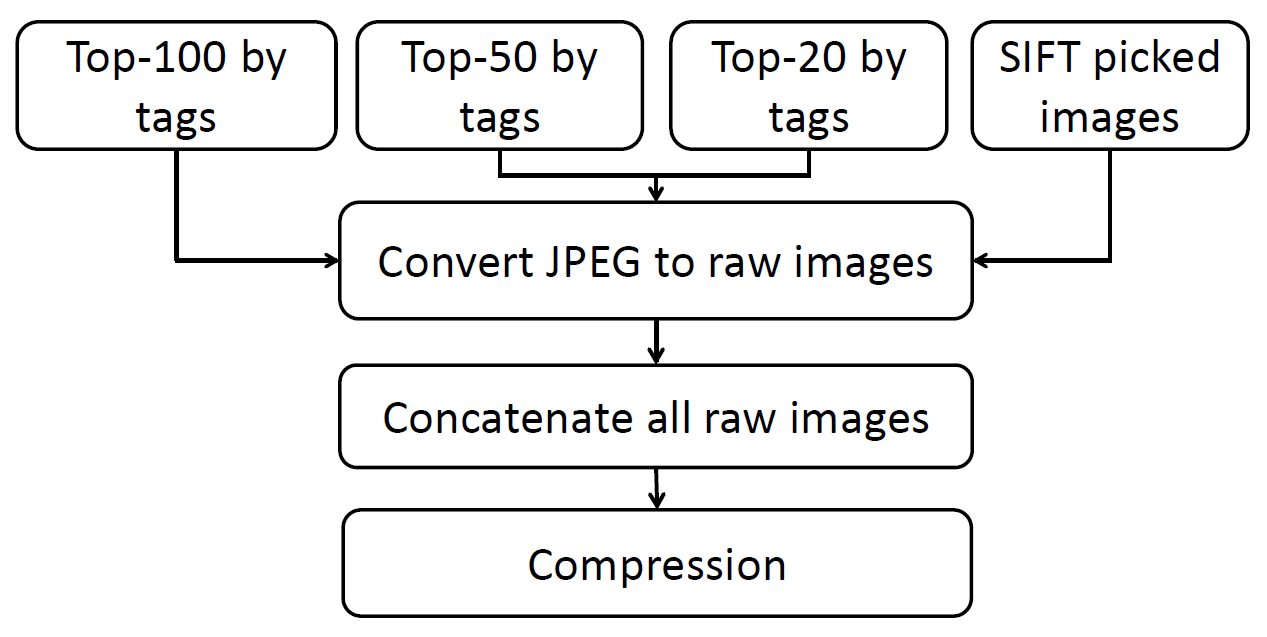}}
  	
   	\caption{Work flow in the experiment: how photo images are processed to assess the compression results}
\label{fig:flow}
\end{figure}

\begin{figure}[h!]
\centering
 \centerline{\includegraphics[width=0.5\textwidth, height=1.5in]{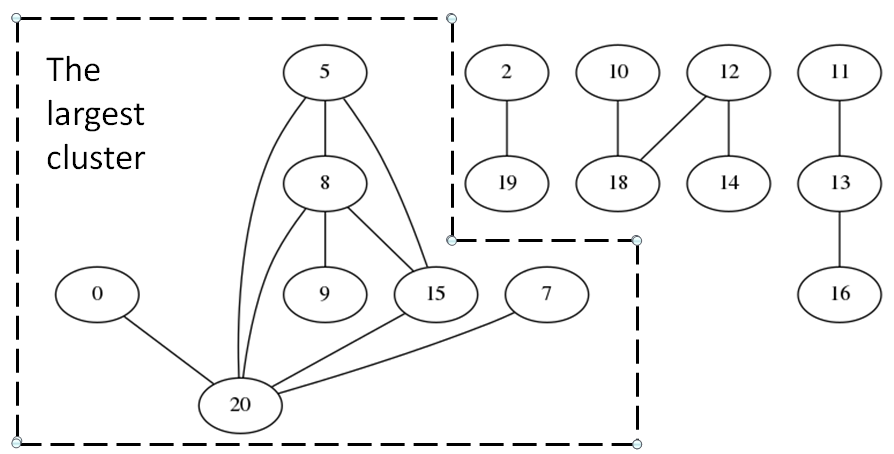}}
  	
   	\caption{An example of choosing the largest cluster from the SIFT results}
\label{fig:sift}
\end{figure}

All the photos downloaded from Flickr are in JPEG format. As shown in Fig.~\ref{fig:flow}, first all JPEG images are decompressed into PNM-format raw image files using djpeg ~\cite{djpeg} for all subgroups. Then, we concatenate all raw image files into one single big file. Finally, we apply two compression tools rzip v2.1 ~\cite{kolivas2008long} and 7-Zip v15.14 ~\cite{pavlov20137zip} to perform the compression. The compressed files are in .rzip and .7z formats respectively. By doing so, we are able to check the inter-file compressibility by leveraging the intra-file optimization in the compression tools. We define compression factor (CF) below as the size of the original file $S_{old}$ divided by the size of new (compressed) file $S_{new}$. The higher the CF is, the better compression result is obtained.  Since we run this experiment to exploit the potential of compressibility, we do not consider the decompression phase in which photo images are to be restored from the single file. The execution time of concatenation and compression is not examined either.
\begin{displaymath}
CF = \frac{S_{old}}{S_{new}}\qquad
\end{displaymath}
For comparison purpose, we use VLFeat v0.9.20 ~\cite{vedaldi08vlfeat} to extract all SIFT local features from Top-100 image groups. Then, we use code from ~\cite{solem2012programming} to compare the features from any two images and get the number of shared ones.  The number of shared features represent the similarity between the two photos. The more features they share, the more similar the two images are. The threshold value for shared features is set to 10 throughout our experiment to eliminate less relevant pairs. Identifying all images that share more than 10 features with each other is a high-dimensional computational problem. To simply the computation, we reduce the problem to finding the cluster with most number of photos. It is a trade-off between the similarity and computation complexity. By doing so, we are able to get a group of photos that are similar more quickly. We visualize the cluster selection process to make it easy to understand. Each image represents a node $n_{i}$ and the group is a set of nodes namely $N = \{n_0,n_1,n_2,n_3,...,n_t\}$ where t = 99. If two images $n_{i}$ and $n_{j}$ share at least ten local features, an edge $e_{ij}$ is established between $n_{i}$ and $n_{j}$. As a result, a diagram like Fig.~\ref{fig:sift} is generated. In Fig.~\ref{fig:sift}, there are four clusters in total. We select the first one as it is the largest cluster with seven members. The other smaller clusters are disregarded. Consequently, the images from the largest cluster are selected as ``SIFT-picked images''. The compression procedure illustrated in Fig.~\ref{fig:flow} is repeated on these images, in additional to Top-100, Top-50 and Top-20 image groups. 

\begin{figure}[h!]
\centering
 \centerline{\includegraphics[width=0.5\textwidth]{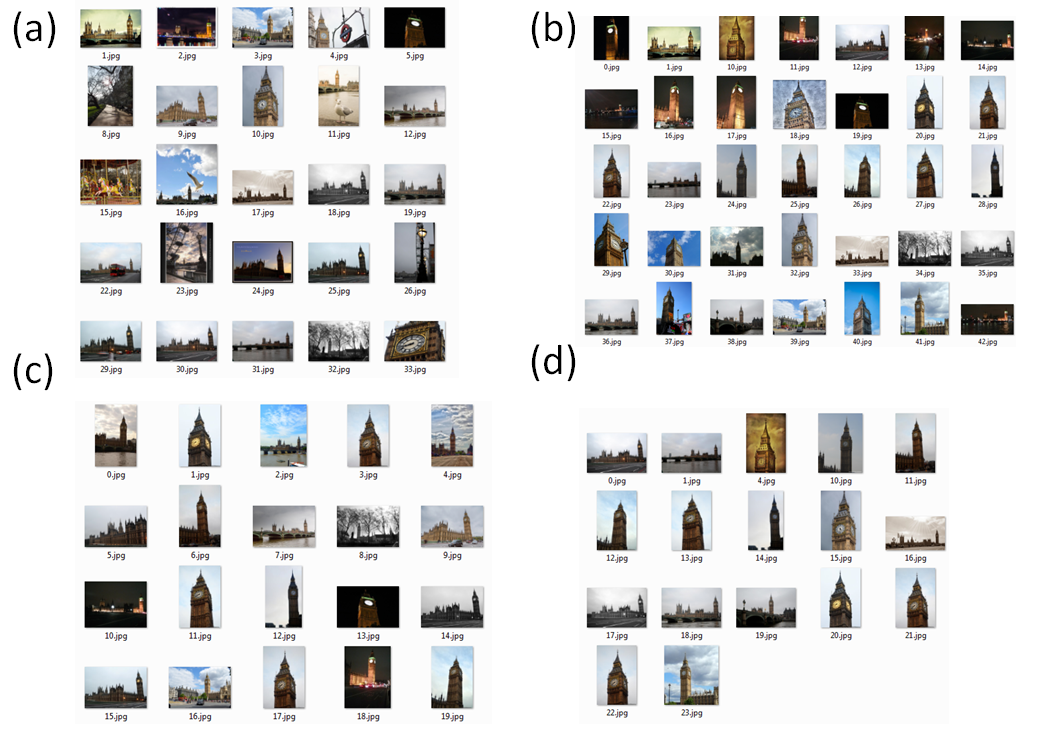}}
  	
   	\caption{Example thumbnails from Top-100, Top-50, Top-20 and SIFT-picked similar photos subgroups with the same tag}
\label{fig:ss1}
\end{figure}

We created 12 groups of photos, each with top 100 photos for the given tags. Then we also create top 50 and top 20 photo groups for comparison purpose. Fig.~\ref{fig:ss1} shows some thumbnails: (a) from Top-100; (b) from Top-50; (c) from Top-20;(d) from SIFT-picked Top-100 images for the tag ``thebigben''. We denote photo group $g_{i}$ where $i = \{1,2...12\}$. We also create two mixed groups from the 1,200 photos by random selection, named m1 and m2. Two more groups are then create by random download from Flickr, named r1 and r2 respectively. 
\section{Evaluation}

All evaluation results are obtained from a workstation equipped with one Intel Core i5 processor with 8GB RAM and 2TB disk space. Our data set includes 1,600 photos with the total size of 817MB acquired using the methods described in Section 3. 

\begin{figure*}[t]
\centering
\subfigure[Results with rzip]{
			\includegraphics[width=0.9\textwidth, height=0.19\textwidth]{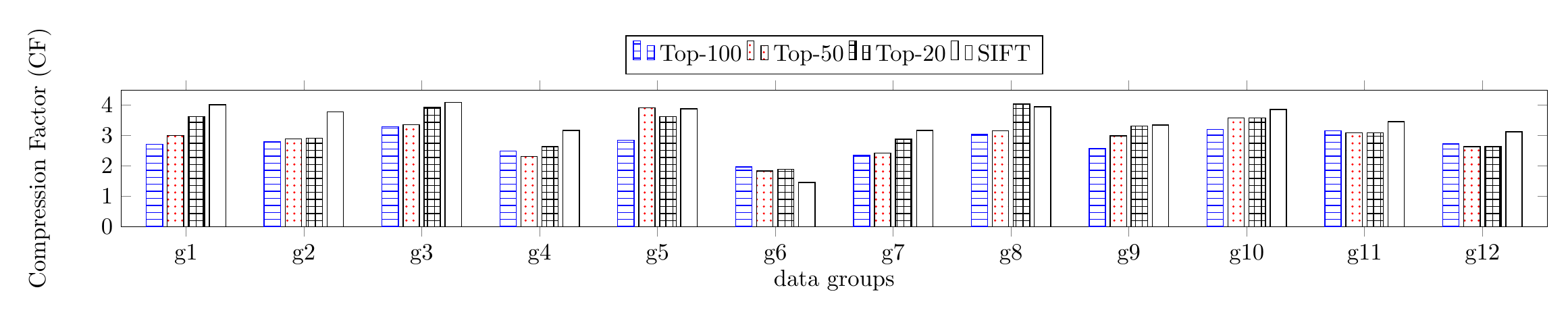}
      }
 	    \hspace{0.05\textwidth}
\subfigure[Results with 7z]{
			\includegraphics[width=0.9\textwidth, height=0.19\textwidth]{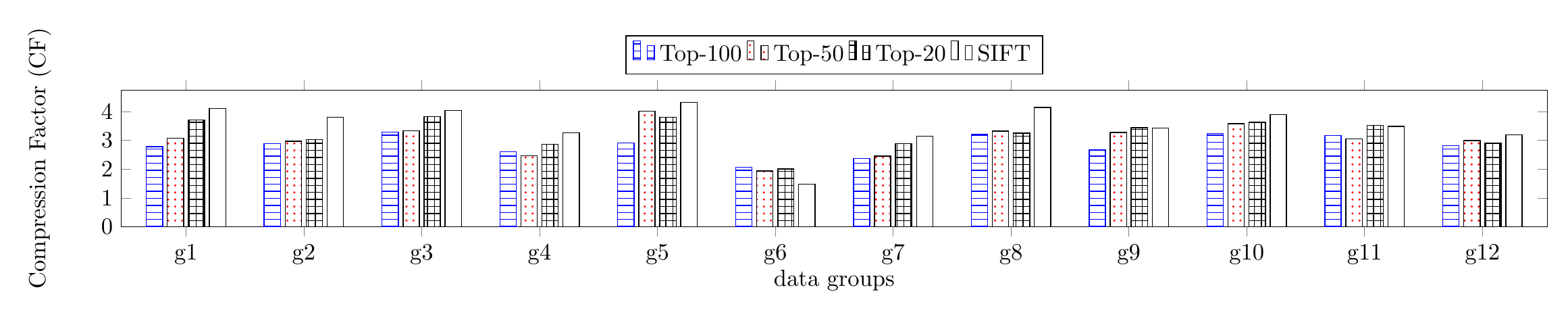}
      }
  	  	\caption{Comparison of compression ratio for Flickr tagged Top-100, Top-50, Top-20 and SIFT-picked subgroups. g1 to g12 represent the tagged groups. }
\label{fig:results_1}
\end{figure*}

The CF for all twelve groups are listed in Fig.~\ref{fig:results_1}. First, we examine the results between Top-100 and Top-50 subgroups. Among the twelve groups, four groups (g1, g5, g9 and g10) see higher CF with Top-50 than Top-100. For other groups, the CF results are either very close between the Top-50 and Top-100 or lower CF is obtained on Top-50. Results from two compressors are quite consistent. Then, we look at the results between Top-100 and Top-20 subgroups. This time, more than half of the groups see a significant higher CF with Top-20, about 10\% in average and up to 26\%. Only one group g6 gets a lower CF with Top-20 subgroup. For the rest, almost equal CF results are observed. For both compression tools, SIFT-picked photos yield a higher CF than Top-100, Top-50 and Top-20 for ten groups out of twelve (about an additional 10\% compared to Top-20) and an almost equal CF for the rest two. Overall, the results from two compression tools are quite close. The exception is g8 for which rzip achieves much greater CF with Top-20. With these results, we can see that on a statistical basis, CF with SIFT-picked is better than Top-20, which is better than Top-50, followed by Top-100. The more relevant (similar) the images are, the higher CF is expected. 

\begin{figure}[h!]
\centering
 \centerline{\includegraphics[width=0.45\textwidth, height=1.3in]{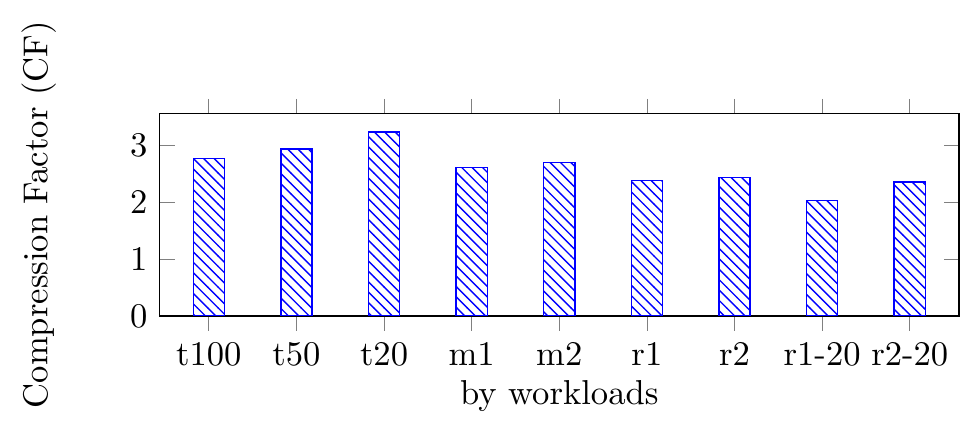}}
  	
   	\caption{Mean CF from Top-100, Top-50 and Top-20 subgroups vs. mixed and random data sets using rzip. t100 stands for mean from Top-100; t50 from Top-50 and t20from Top-20. r1, r2 represent the randomly downloaded photo groups; m1, m2 represent two mixed photos from g1 through g12 repository}
\label{fig:mean}
\end{figure}

It is interesting to analyze what factors may impact the correlation of tag and CF. We find where CF improvements are more distinctive, such as g1 (thebigben), g3 (tajmahal), g7 (milfordsound) and g8 (oriental pearl), the relevant objects are symbolic and easy to be identified. Multiple tags do not make significant difference. In contrast, g6 (pizza, pepperoni) does not have a concrete pattern. And moreover, the SIFT-picked image set only includes two images for g6. There are only two images sharing at least ten SIFT local features, reflecting the diversity of the images in g6. g6 is regarded as an anomaly. According to Fig.~\ref{fig:mean} , the mean CF for all Top-100 subgroups (rzip) is 2.76 while the CF of m1 and m2 are slightly lower (2.65 and 2.74). Mixed images from the same pool yield lower CF as the relevance of the group goes down. For random groups, we actually see a different pattern when the group contains fewer photos (20 photos vs. 100 ones). We explain this with compression tool mechanism: when data is randomly organized, the CF is determined by the hit rate of identical contents in the compressor dictionary. The bigger the data pool is , the more likely the new incoming data gets a hit, thus yield a higher CF. Based on the results above, we believe Flickr tags are helping users to get more relevant images. 

\begin{figure}[h!]
\centering
 \centerline{\includegraphics[width=0.42\textwidth, height=1.3in]{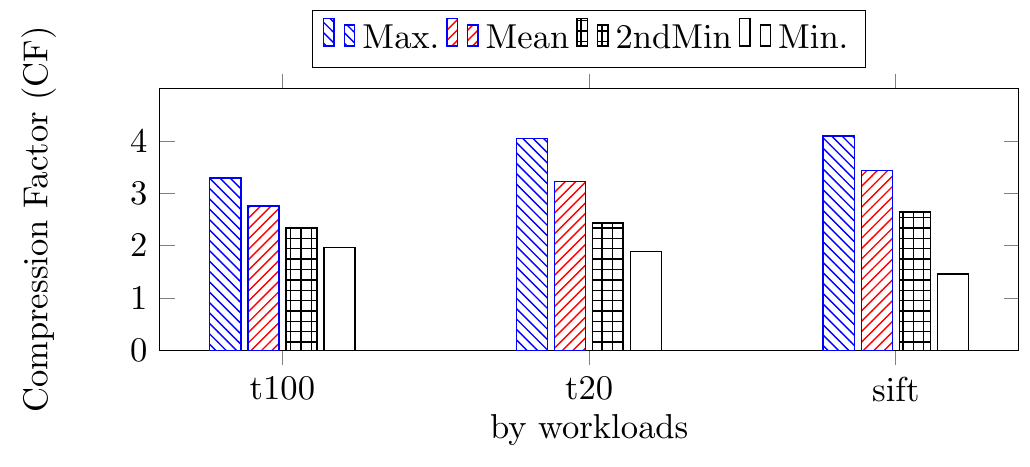}}
  	
   	\caption{The maximum, mean, the 2nd minimum and the minimum of CF, from Top-100, Top-20 and SIFT-picked subgroups using rzip. This is a statistical view of the storage efficiency of these group. }
\label{fig:siftres}
\end{figure}
Fig.~\ref{fig:siftres} shows the maximum, mean and minimum CF across the twelve groups we test with Top-100, Top-20 and SIFT-picked selection. As previously discussed, g6 is a anomaly and all its results represent the minimum of the three selection groups. So we add the 2nd minimum CF data to gain more insights. We find that statistically, the CF from SIFT-picked is 10-15\% better than CF from Top-20, which outperforms CF from Top-100 by around 15\%. By checking thumbnails illustrated in Fig.~\ref{fig:ss1}, we find that both SIFT-picked and Top-20 can help to gather more similar photos than Top-100 does. Extracting SIFT local features from 100 photos takes 15-20 minutes while sorting photos by tag relevance almost takes no computing time on the client side. SIFT approach is more accurate than tags at the cost of extra computation. The huge amount of ``sorting'' work has been accomplished by the users when photo are uploaded, or by the platform back-end program using unknown algorithms. Therefore, tags can be used as an efficient similarity detection, grouping and data placement approach.

In summary, we found correlation between photo tags and compressibility which helps to improve the storage spatial efficiency. A few limitations for tag selection are discovered. When the tags are referring to a specific and distinctive object, the correlation is higher. The mechanism of the tagging algorithm may also affect the correlation levels. It is a matter of how accurate the pre-processing can be. The comparison with SIFT local feature extraction shows that there is enough space for improvements. Ideally, the tag relevance may achieve storage efficiency results close to SIFT approach. More importantly, what we have discussed is complementary to what JPEG has done for digital image compression.

\section{Related Work}

Recently, to embrace the big data era, research community has shifted the focus from general storage efficiency techniques to application and data-aware specialized methods with some pre-processing capabilities. For example, some exploited the separation of metadata from data in tar files ~\cite{lin2015metadata}. By moving metadata to different locations of the file, the deduplication ratio is improved significantly. Conventional wisdom states that video data is difficult to be deduplicated. In ~\cite{dewakar2015storage}, variations such as captions, resolutions, web optimization are evaluated with different deduplication techniques. The results show that with pre-processing, video files can be effectively deduplicated. In addition, migratory compression ~\cite{lin2014migratory} has been proposed to reorder the binary sections before feeding data to compression tools to achieve better intra-file compressibility with trade-off in performance and restoring efforts. A recent study ~\cite{shi2014photo} has utilized local features rather than individual pixel values to analyze the similarity between photos from the same album, to achieve better compression results. 
\section{Conclusion and Future Work}

In this paper, we have employed the data sets and tagging system from Flickr for the empirical study. We observed storage efficiency results from two content-based similarity detection approaches for raw digital images. The results showed that with the help of similarity, the compression factor can be improved significantly, by up to 26\%. The insights obtained from the study may help direct the future system design. Our future work includes measuring and optimizing time efficiency of the aforementioned similarity detection approaches. We also expect new storage system design to be developed to pre-process raw images and utilize inter-file content-based similarity, which achieves greater storage efficiency.
%



{\footnotesize \bibliographystyle{acm}
\bibliography{ref}}

\end{document}